\documentclass[doublecol]{epl2}

\usepackage{amsmath} 
\usepackage{graphicx}
\usepackage{dcolumn} 
\usepackage{bm}
\usepackage{epsf}
\usepackage{epsfig}

\title{Quantum discord dynamics in structured reservoirs}
\shorttitle{Quantum discord dynamics in structured reservoirs} 

\author{Z. -K. Su \and S. -J. Jiang\thanks{E-mail: \email{stsjsj@mail.sysu.edu.cn}}}
\shortauthor{Z. -K. Su \etal}

\institute{
State Key Laboratory of Optoelectronic Materials and Technologies, Sun Yat-Sen University,

Guangzhou 510275, China
}

\abstract{
The non-Markovian master equations are derived to study quantum discord dynamics of two qubits coupled to a common reservoir and two independent reservoirs, respectively. We compare the dynamics under different parameters, such as reservoir spectra and resonant parameters, at high temperature and at zero temperature. The results indicate that the dynamics at these two extreme temperatures share similar characters, as well as differences.}

\begin{document}

\maketitle

All quantum systems interact with their surrounding environments. The interactions lead to dissipation and decoherence due to a flow of information between the system and the environment~\cite{book1,book2}. Fortunately, reservoir engineering techniques can suppress decoherence in laboratories and thereby make the system avoid suffering from the environment. Engineered reservoirs arise from many physical situations, including photonic crystals~\cite{photonic.crystal}, controllable Ohmic-like environments~\cite{ohmic.like}, as well as optical and microwave cavities~\cite{optical.microwave}. Recent reservoir engineering techniques~\cite{photonic.crystal,ohmic.like,optical.microwave,addition.techniques} aim to alter the dynamics of dissipation and decoherence in an open quantum system by modifying characteristics of the environment, such as reservoir temperatures, reservoir spectra and resonant parameters. Understanding which type of environment leads to faster or slower decoherence dynamics is essential in the choice of the physical system for implementing realistic quantum devices such as a quantum computer. Therefore, it is of fundamental and practical importance to study the decoherence dynamics of a system in structured reservoirs.

Decoherence can be viewed as the loss of nonclassical correlation of a system. Entanglement and quantum discord are two widely used measures of nonclassical correlation. Entanglement receives much attention since it plays a crucial role in quantum information processing. However, as it is discoved that the unentangled states can also have nonclassical~\cite{almost} correlation and that the use of such states can improve performance in some computational tasks~\cite{speedup}, quantum discord is proposed~\cite{zurek.qd} and it is defined as distinction between quantum and classical aspects of correlation in a composite quantum state. Entanglement and quantum discord can behave very differently under certain reservoirs. For example, under Markovian environments~\cite{024103} quantum discord is more robust than the entanglement against decoherence; under independent~\cite{014101} or common~\cite{052107} zero-temperature non-Markovian reservoirs, quantum discord presents sudden changes~\cite{044102} while entanglement shows oscillations with or without suddden death. Experimentally, the above distinguished behaviors have been obtained~\cite{7042328}. Recent results suggest that a change in quantum discord could be used as an indicator of failure of local operations and classical communications~\cite{010301(R)}. Besides, the quantum thermodynamics efficiency of a photo-Carnot engine in terms of quantum discord of an atomic pair can exceed its classical value~\cite{50003}. In this letter, we employ quantum discord to characterize the quantum correlation present in a two-qubit system. We consider these two qubits subject to two different situations, the independent and common environments, and aim to find out under which situation coherence can be preserved longer in high-temperature and in low-temperature regions, respectively. With Einstein convention sum being adopted, the Hamiltonian for the case of independent reservoirs reads $(\hbar=1)$

\begin{equation}
\label{eq.independenthami}
\emph{H}_i=\omega_{a}\sigma^{i}_{+}\sigma^{i}_{-}+\sum_{k}\omega^{i}_{k}a^{i^{\dag}}_{k}a^{i}_{k}+(\sigma^{i}_{+}\textbf{\emph{B}}^{i}+\sigma^{i}_{-}\textbf{\emph{B}}^{i^{\dag}})
,
\end{equation}
and the Hamiltonian for the case of a common reservoir takes the form
\begin{equation}
\label{eq.commonhami}
\emph{H}_c=\omega_{a}\sigma^{i}_{+}\sigma^{i}_{-}+\sum_{k}\omega_{k}a^{\dag}_{k}a_{k}+(\sigma^{i}_{+}\textbf{\emph{B}}+\sigma^{i}_{-}\textbf{\emph{B}}^{{\dag}})
,
\end{equation}
where $\omega_{a}$ is the transition frequency of the qubits, $\sigma^{i}_{\pm}$ are the system raising and lowering operators of the $\emph{i}$th qubit, and $\textbf{\emph{B}}^{i}=\sum_{k}g^{i}_{k}a^{i}_{k}$ ($\textbf{\emph{B}}=\sum_{k}g_{k}a_{k}$) with $g^{i}_{k}$ ($g_{k}$) being the coupling constants. In the limit of a continuum of reservoir modes $\sum_{k}|g_{k}|^{2}\rightarrow\int\emph{d}\omega\emph{J}(\omega)$, where $\emph{J}(\omega)$ is the spectral density function, charactering the reservoir spectrum. Here the index $\emph{k}$ labels the reservoir field modes with frequencies $\omega^{i}_{k}$ ($\omega_{k}$), and $a^{i^{\dag}}_{k}$ and $a^{i}_{k}$ ($a^{\dag}_{k}$ and $a_{k}$) are their usual creation and annihilation operators, respectively.

Let us now proceed to the master equation for this two-qubit system using the method suggested in~\cite{master.equation}. In the weak coupling limit, assuming an initially factorized state and
a thermal reservoir, we obtain from eq.~(\ref{eq.independenthami}) a secularly approximated non-Markovian master equation in the interaction picture

\begin{align}
\label{independentME}
\frac{d\rho}{dt}=&-4\kappa_1\rho-i\kappa_2J_0\rho\nonumber\\
&+2(\kappa_1+\mu_1)K_{-}\rho+2(\kappa_1-\mu_1)K_{+}\rho-4\mu_1K_{0}\rho\,
\end{align}

and stating from eq.~(\ref{eq.commonhami}), we have

\begin{align}
\label{commonME}
\frac{d\rho}{dt}=&-4\kappa_1\rho-2\kappa_1J_1\rho-2i\mu_2J_2\rho\nonumber\\
&+2(\kappa_1+\mu_1)J_{-}\rho+2(\kappa_1-\mu_1)J_{+}\rho-i\kappa_2J_0\rho\nonumber\\
&+2(\kappa_1+\mu_1)K_{-}\rho+2(\kappa_1-\mu_1)K_{+}\rho-4\mu_1K_{0}\rho\,
\end{align}

where $J_{0,1,2,\pm}$ and $K_{0,\pm}$ are superoperators defined as
\begin{equation}
\label{eq.J0}
    J_0\rho=\sigma_{z}^{(1)}\rho+\sigma_{z}^{(2)}\rho-\rho\sigma_{z}^{(1)}-\rho\sigma_{z}^{(2)},
\end{equation}
\begin{equation}
\label{eq.J1}
    J_1\rho=\sigma_{-}^{(1)}\sigma_{+}^{(2)}\rho+\sigma_{+}^{(1)}\sigma_{-}^{(2)}\rho+\rho\sigma_{-}^{(1)}\sigma_{+}^{(2)}+\rho\sigma_{+}^{(1)}\sigma_{-}^{(2)},
\end{equation}
\begin{equation}
\label{eq.J2}
    J_2\rho=\sigma_{+}^{(1)}\sigma_{-}^{(2)}\rho+\sigma_{+}^{(2)}\sigma_{-}^{(1)}\rho-\rho\sigma_{+}^{(1)}\sigma_{-}^{(2)}-\rho\sigma_{+}^{(2)}\sigma_{-}^{(1)},
\end{equation}
\begin{equation}
\label{eq.J-}
    J_{-}\rho=\sigma_{-}^{(1)}\rho\sigma_{+}^{(2)}+\sigma_{-}^{(2)}\rho\sigma_{+}^{(1)},
\end{equation}
\begin{equation}
\label{eq.J+}
    J_{+}\rho=\sigma_{+}^{(1)}\rho\sigma_{-}^{(2)}+\sigma_{+}^{(2)}\rho\sigma_{-}^{(1)},
\end{equation}
\begin{equation}
\label{eq.K-}
    K_{-}\rho=\sigma_{-}^{(1)}\rho\sigma_{+}^{(1)}+\sigma_{-}^{(2)}\rho\sigma_{+}^{(2)},
\end{equation}
\begin{equation}
\label{eq.K+}
    K_{+}\rho=\sigma_{+}^{(1)}\rho\sigma_{-}^{(1)}+\sigma_{+}^{(2)}\rho\sigma_{-}^{(2)},
\end{equation}
\begin{align}
\label{eq.K0}
K_0\rho=&\frac{\sigma_{+}^{(1)}\sigma_{-}^{(1)}\rho+\rho\sigma_{+}^{(1)}\sigma_{-}^{(1)}-\rho}{2}\nonumber\\
&+\frac{\sigma_{+}^{(2)}\sigma_{-}^{(2)}\rho+\rho\sigma_{+}^{(2)}\sigma_{-}^{(2)}-\rho}{2},
\end{align}
and the time dependent coefficients $\kappa_i(t)$ and $\mu_i(t)$ $(i=1,2)$ can be expressed as power series in the system reservoir coupling constant $\alpha$ [to be defined in eq.~(\ref{eq.ohmic})]. For weak coupling (i.e. when $\alpha\ll1$), one can stop the expansion to the second order and obtain analytic solutions for these coefficients. In thermal equilibrium, these coefficients read
\begin{equation}\label{eq.kappa1}
    \kappa_1(t)=\frac{1}{2}\int^{t}_{0}d\tau\int d\omega J(\omega)(1+2N(\omega))\cos(\omega-\omega_{a})\tau,
\end{equation}
\begin{equation}\label{eq.kappa2}
    \kappa_2(t)=\frac{1}{2}\int^{t}_{0}d\tau\int d\omega J(\omega)(1+2N(\omega))\sin(\omega-\omega_{a})\tau,
\end{equation}
\begin{equation}\label{eq.mu1}
    \mu_1(t)=\frac{1}{2}\int^{t}_{0}d\tau\int d\omega J(\omega)\cos(\omega-\omega_{a})\tau,
\end{equation}
\begin{equation}\label{eq.mu2}
    \mu_2(t)=\frac{1}{2}\int^{t}_{0}d\tau\int d\omega J(\omega)\sin(\omega-\omega_{a})\tau,
\end{equation}
where $N(\omega)=(e^{\omega/k_{B}T}-1)^{-1}$ is the average number distribution of the reservoir thermal excitation with frequency $\omega$ at the initial time, $k_{B}$ the Boltzmann constant, and $T$ the reservoir temperature. A closed form for the expressions of the time-dependent coefficients can be obtained in the high-temperature and zero-temperature limits, i.e., for $1+2N(\omega)\approx 2k_{B}T/\omega$ and $1+2N(\omega)\approx 1$, respectively. The solutions to the master equations ~(\ref{independentME}) and ~(\ref{commonME}) can be obtained in the supplementary material.

All the dynamics characters of the system are determined by the time dependent coefficients, $\kappa_2$, $\mu_2$, $\kappa_1$ and $\mu_1$, in the master equations ~(\ref{independentME}) and ~(\ref{commonME}). The difference between eq.~(\ref{independentME}) and eq.~(\ref{commonME}) is that the latter induces superoperators $J_{1,2,\pm}$. From eq.~(\ref{eq.J0}) $\sim$  eq.~(\ref{eq.K0}), we know that $J_{1,2,\pm}$ are caused by the coupling between qubits, while $J_0$ and $K_{0,\pm}$ are introduced by each independent qubit. Now two questions come up; do superoperators $J_{1,2,\pm}$ speed up or slow down the decoherence? does the role of $J_{1,2,\pm}$ vary between high and low temperature region, between resonant and off-resonant region, between different spectra of reservoir? We will answer these questions in the following.

\begin{figure*}
\vspace{-1cm}
\includegraphics[width=18cm]{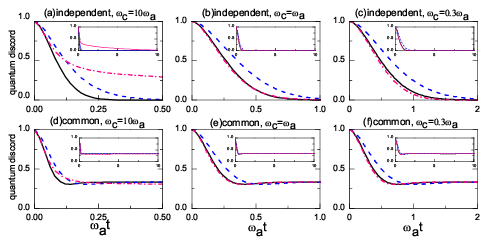}
\vspace{-1.5cm}
\caption{(Colors online) Time evolutions of the quantum discord of the two qubits with an initial Bell state $|\Psi\rangle=(|e_1g_2\rangle+|g_1e_2\rangle)/\sqrt{2}$. (a) $\omega_c=10\omega_a$ for independent reservoirs, (b) $\omega_c=\omega_a$ for independent reservoirs, (c) $\omega_c=0.3\omega_a$ for independent reservoirs, (d) $\omega_c=10\omega_a$ for a common reservoir, (e) $\omega_c=\omega_a$ for a common reservoir, (f) $\omega_c=0.3\omega_a$ for a common reservoir. The black solid line is for sub-Ohmic reservoir, blue dashed for Ohmic reservoir and pink dash-dotted for super-Ohmic reservoir. The insets show the same plots but with much larger values of horizontal axis $\omega_at$. We have set $k_BT/(\hbar\omega_a)=100$.
\label{fig1}}
\end{figure*}

In order to compare the dynamics for different types of reservoirs, we consider a class of spectral densities. The spectral densities we examine are of the form
\begin{equation}\label{eq.ohmic}
    J(\omega)=\alpha^{2}\omega_c^{1-s}\omega^{s}e^{-\omega/\omega_c},
\end{equation}
which is classified as sub-Ohmic if $0<s<1$, Ohmic if $s=1$, and super-Ohmic if $s>1$. We consider three examples with $s=1/2,1$ and $3$ corresponding to sub-Ohmic, Ohmic, and super-Ohmic spectral densities, respectively. The parameter $\omega_c$ is the cutoff of high frequency of reservoir and is connected to the reservoir correlation time $\tau_c\approx\omega_c^{-1}$. On the other hand, the dimensionless coupling constant $\alpha$ is related to the relaxation time scale $\tau_r\approx\alpha^{-2}$, over which the state of the system changes, in the Markovian limit of flat spectrum. Generally, the dynamics of a quantum system comprises three different dynamical effects occurring at three different respective time scales. Firstly, the dynamics of decoherence occurs at a time scale of the order of the relaxation time scale $\tau_r$, which is defined by the properties of the reservoir. Secondly, nonsecular terms cause oscillations, occurring over a typical time scale $\tau_a\approx\omega_a^{-1}$ of the system. Finally, the non-Markovian memory effects happen for a time shorter than or of the order of the reservoir correlation time scale $\tau_c$. In the present work, we focus on both short-time and long-time evolution at the secular regime characterized by the condition $\tau_a\ll\tau_r$~\cite{book1}. In all the plots presented below, we set the coupling constant $\alpha^2=0.01\omega_a$, which satisfies the secular approximation condition $\tau_a\ll\tau_r$.

Now let us consider a bipartite state $\rho_{AB}$ to review the definition of quantum discord~\cite{zurek.qd,vedral.qd}. The idea of quantum discord grows out of the fact that the quantum mutual information of the state $\rho_{AB}$ may be determined in two nonequivalent ways. The first is obtained by $\mathcal{I}(\rho_{AB})=S(\rho_{A})+S(\rho_{B})-S(\rho_{AB})$, where $S(\rho)=-Tr[\rho log \rho]$ is the von Neumann entropy of the state $\rho$ and $\rho_{A(B)}=Tr_{B(A)}[\rho_{AB}]$ are the partial traces over the two subsystems. On the other hand, one may acquire the quantum mutual information based on condition entropy, that is, $J_A=max_{\{\prod_k\}}[S(\rho_A)-S(\rho_{AB}|\{\prod_k\})]$, where the maximum is taken over the set of position-operator-valued measurement (POVM) $\{\prod_k\}$ in partition B, $S(\rho_{AB}|\{\prod_k\})=\sum_kp_kS(\rho_k)$ is the quantum condition entropy, $\rho_k=(I\otimes \prod_k)\rho_{AB}(I\otimes \prod_k)/Tr(I\otimes \prod_k)\rho_{AB}(I\otimes \prod_k)$ is the conditional density operator corresponding to the outcome labeled by $k$, and $p_k=Tr(I\otimes \prod_k)\rho_{ab}(I\otimes \prod_k)$. Here \emph{I} is the identity operator on subsystem $A$. Finally, the quantum $A$ discord is defined in terms of the mismatch $Q(\rho_{AB})=\mathcal{I}(\rho_{AB})-J_A(\rho_{AB})$. Similarly, one can also define the $B$ discord through the entropy of conditional states of system $B$. In the following, we will employ the measure of quantum discord to study under which situations the coherence can be maintained longer in high and low temperature region, respectively. 

\begin{figure*}
\vspace{-1cm}
\includegraphics[width=18cm]{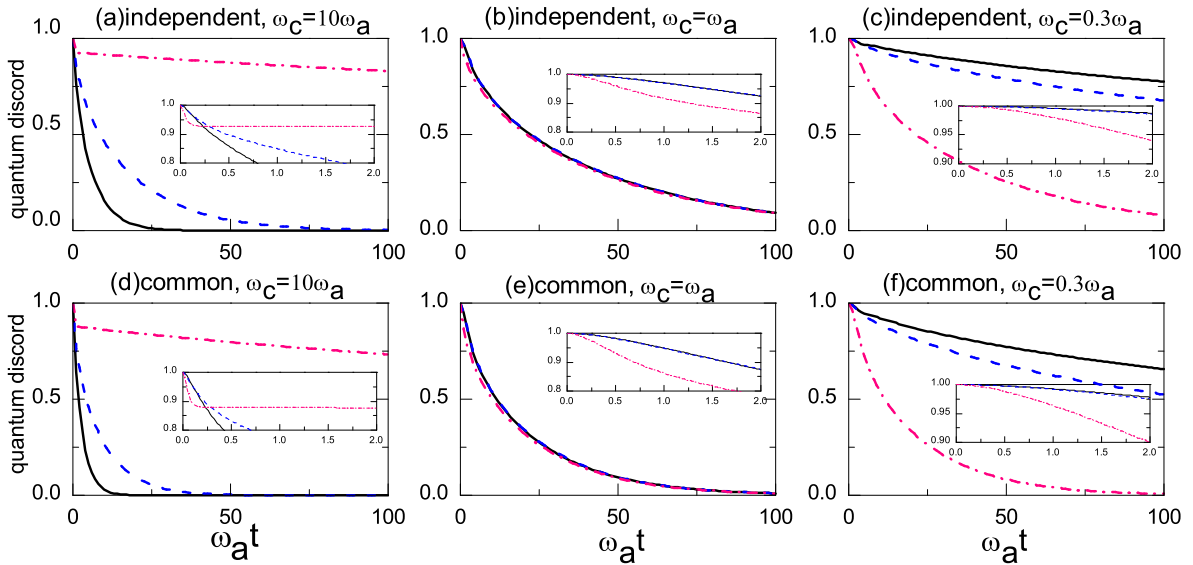}
\vspace{-1.5cm}
\caption{(Colors online) Time evolutions of the quantum discord of the two qubits with an initial Bell state $|\Psi\rangle=(|e_1g_2\rangle+|g_1e_2\rangle)/\sqrt{2}$. (a) $\omega_c=10\omega_a$ for independent reservoirs, (b) $\omega_c=\omega_a$ for independent reservoirs, (c) $\omega_c=0.3\omega_a$ for independent reservoirs, (d) $\omega_c=10\omega_a$ for a common reservoir, (e) $\omega_c=\omega_a$ for a common reservoir, (f) $\omega_c=0.3\omega_a$ for a common reservoir. The black solid line is for sub-Ohmic reservoir, blue dashed for Ohmic reservoir and pink dash-dotted for super-Ohmic reservoir. The insets show the same plots but with much smaller values of vertical axis $quantum$ $discord$ and horizontal axis $\omega_at$. We have set $T=0$.
\label{fig2}}
\end{figure*}

We begin addressing the high-temperature region, where we choose a temperature such that $k_BT=100\omega_a$. Fig.~\ref{fig1} shows us the time evolution of the quantum discord of the two qubits with an initial Bell state $|\Psi\rangle=(|e_1g_2\rangle+|g_1e_2\rangle)/\sqrt{2}$, coupled to a common reservoir or different individual reservoirs at high temperature, for three different values of resonance parameters, $\omega_c=10\omega_a$, $\omega_c=\omega_a$ and $\omega_c=0.3\omega_a$. We first compare quantum discord evolution under a common reservoir with that under independent reservoirs. As shown in fig.~\ref{fig1}(c) and (f), the quantum discord is preserved in a steady state in the case of a common reservoir, while in the case of independent reservoirs it decays exponentially with time. This phenomenon is due to the difference between eq.~(\ref{independentME}) and eq.~(\ref{commonME}), that is, the correlated shifted frequency induced by $J_2$ and decay rate caused by $J_{1,\pm}$ between the two qubits. Note that the behavior agrees with the results of~\cite{steady.state} where the entanglement dynamics for two oscillators or two qubits in the same environment was discussed. The model that a two-qubit system is coupled to a common bath will attract much attention, because under this model one may preserve quantum coherence at high temperature, which makes it a major distinct feature with respect to the case of separate baths. Similar results can be obtained by comparing fig.~\ref{fig1}(a) with (d) and (b) with (e). We next compare the three different Ohmic-like reservoirs and find out which one induces the slowest decoherence. From fig.~\ref{fig1}, we can see that the Ohmic reservoir induces the slowest decoherence, while for the super-Ohmic and sub-Ohmic reservoirs the discord decay in a very similar manner, both faster than the Ohmic case, in short-time region. Therefore, if one is able to modify the natural reservoir spectrum into an Ohmic form, one would slow down decoherence with respect to the sub-Ohmic and super-Ohmic ones. This conclusion is consistent with the results found in~\cite{a.oscillator.1} and~\cite{two.fields.1}, where a quantum harmonic oscillator and two cavity fields are investigated, respectively. However, it should be mentioned that fig.~\ref{fig1}(a) also indicates that the super-Ohmic independent reservoirs induce the slowest decoherence in long-time region. Finally, the effect of resonance parameter will be reviewed. Changing this parameter corresponds to shifting the qubit frequency with respect to the reservoir spectrum. This allows us to control the effective coupling between the system and the environment. For $p\ll1$ $(p=\omega_c/\omega_a)$ the system is off-resonant with respect to the peak of the reservoir spectrum. Accordingly, for $p\gg1$ the system is resonant with respect to the peak of the reservoir spectrum. From previous results~\cite{previous.results}, we expect to see different dynamics in the $p\ll1$ and $p\gg1$ regimes. Comparing fig.~\ref{fig1}(a) with (c), (d) with (f), we find that, whatever type of reservoir the system is under, for $p\gg1$ the decoherence process is faster than that for $p\ll1$ in short-time region. This dues to the overlap between the frequency of the system and the reservoir spectrum in the resonant case. But in long-time region the decoherence process is significantly slower for $p\gg1$ than for $p\ll1$ under the influence of independent super-Ohmic reservoirs, as shown in fig.~\ref{fig1}(a) and (c). Thus in the $p\gg1$ case the effective coupling of the system to the reservoir is obviously weaker than that in the $p\ll1$ case for super-Ohmic reservoir in long-time region. In summary, the results show that (i) A common reservoir forms a steady state at high temperature, which is superior to the case of two separate reservoirs; (ii)For Ohmic reservoir the system shows the longest quantum coherence in each situation except for the situation of independent reservoir in resonant region, where for super-Ohmic reservoir it appears to have the weakest effect on the system in long-time region; (iii) Under sub-Ohmic and Ohmic reservoirs, the coherence will be maintained longer in the off-resonant region, while for super-Ohmic reservoir we should consider it for two cases. In case of short-time region, quantum correlation can be preserved longer in the off-resonant region, however, in case of long-time region this correlation can be protected for a longer time.

\begin{figure*}
\vspace{-1cm}
\includegraphics[width=18cm]{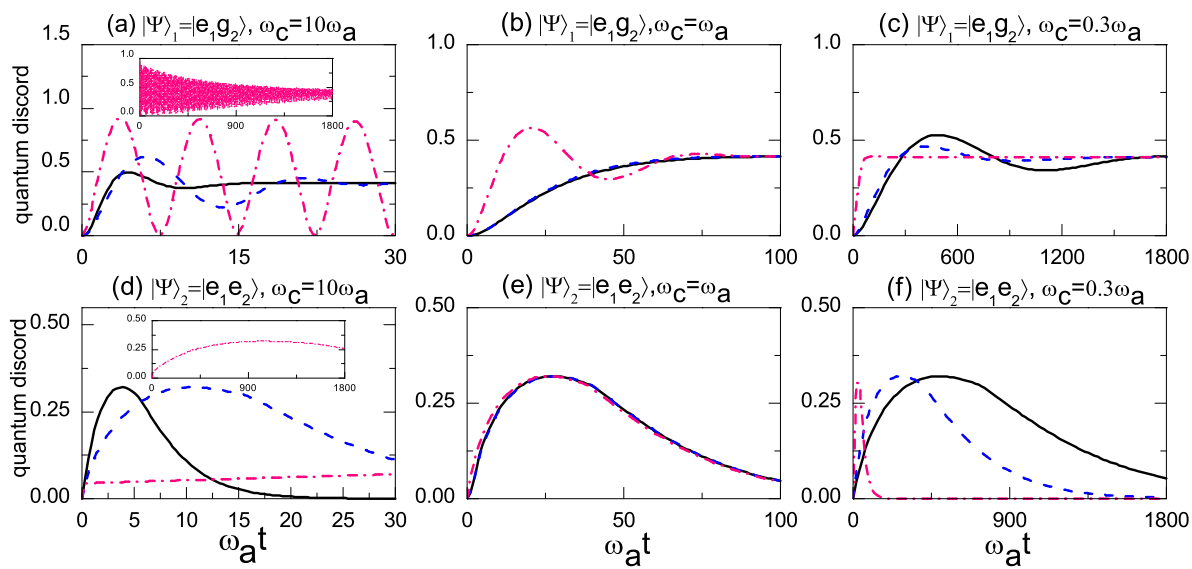}
\vspace{-1.5cm}
\caption{(Colors online) Time evolutions of the quantum discord of the two qubits with an initial separable state $|\Psi\rangle_1=|e_1g_2\rangle$ or $|\Psi\rangle_2=|e_1e_2\rangle$ for a zero-temperature common reservoir. (a) $\omega_c=10\omega_a$ with $|\Psi\rangle_1$, (b) $\omega_c=\omega_a$ with $|\Psi\rangle_1$, (c) $\omega_c=0.3\omega_a$ with $|\Psi\rangle_1$, (d) $\omega_c=10\omega_a$ with $|\Psi\rangle_2$, (e) $\omega_c=\omega_a$ with $|\Psi\rangle_2$, (f) $\omega_c=0.3\omega_a$ with $|\Psi\rangle_2$. The black solid line is for sub-Ohmic reservoir, blue dashed for Ohmic reservoir and pink dash-dotted for super-Ohmic reservoir. The insets show very-long-time scale dynamics. We have set $T=0$.
\label{fig3}}
\end{figure*}

We now move on to the zero-temperature region, i.e. $T=0$. From previous studies~\cite{zero.temperature} on open quantum system interacting with zero-temperature reservoirs, we expect an apparently slower loss of coherence with respect to the $T\neq0$ case. Nevertheless questions still remain. Does the zero-temperature common reservoir induce a steady state like the high-temperature common reservoir? Does Ohmic environment still show the weakest dissipation-noise effect on the decoherence dynamics? Does the resonant parameter have the same influence as at high temperature? In order to answer these questions, in fig.~\ref{fig2} we plot time evolution with the same parameters as fig.~\ref{fig1} except for temperature. As shown in fig.~\ref{fig2}, the zero-temperature common reservoir not only doesn't induce a steady state, but also speeds up the decoherence rate, compared with the case of independent reservoir, on condition that other parameters are the same. As for the second question, we discuss short-time and long-time region separately. From the insets of fig.~\ref{fig2}, we find that super-Ohmic environment shows the strongest decoherence, while the sub-Ohmic and Ohmic reservoirs decay in a very similar manner, both slower than the super-Ohmic case, in short-time region. By contrast, it is not easy to say which type of spectra induces the slowest decoherence in long-time region, since it dependents on resonant parameters. For example, sub-Ohmic reservoir shows the weakest effect in case $\omega_c=0.3\omega_a$, but in case $\omega_c=10\omega_a$ super-Ohmic reservoir does. Finally, fig.~\ref{fig2} also tells us that the effects of resonant parameters are similar to those at the high temperature, that is, the decoherence process is faster for $p\gg1$ than $p\ll1$ under the influence of Ohmic or sub-Ohmic reservoir, but conversely in case of super-Ohmic reservoir. All in all, the decoherence dynamics in zero-temperature and high-temperature share some indentical characters, but are different from each other as to whether a steady-state will be  produced under a common reservoir. Furthermore, at zero temperature what type of reservoir leads to the weakest decoherence depends on resonant parameters and time regions, while at high temperature Ohmic reservoir always shows the weakest impact on the system. Lastly, the roles of resonant parameters are similar to those in high temperature region.

As well as avoiding decoherence, quantum coherence must also be generated. In fact, it has been reported that two qubits coupled to a single mode~\cite{single.mode} or a common heat bath~\cite{common.bath} can be entangled by purely dissipative dynamics. Here, we focus on creation of quantum discord of two qubits interacting with a zero-temperature common bath and consider the behavior characteristics for varies of parameters. In fig.~\ref{fig3}, we plot the dynamics of this system with different initial states $|\Psi\rangle_1=|e_1g_2\rangle$ and $|\Psi\rangle_2=|e_1e_2\rangle$, respectively. A first look at the plots of~\ref{fig3} shows that quantum discord can be generated while the evolution varies with reservoir spectrum, resonant parameter and initial state. For the initial state $|\Psi\rangle_1=|e_1g_2\rangle$, we find from fig.~\ref{fig3} (a) $\sim$ (c) that the quantum discord increases and approaches to a definite value monotonically. This conclusion is similar to the result found in~\cite{definite.value} where dynamics of bath-induced entanglement was analyzed. And one may also see that the time evolution of quantum discord is changed with the type of spectral densities for the same resonant parameter. For example, as shown in fig.~\ref{fig3} (a), the super-Ohmic environment induces the strongest quantum discord oscillation; next is the Ohmic case; while the sub-Ohmic environment causes a relatively weak quantum discord oscillation. In addition, for a fixed type of spectral density, the behavior of quantum discord varies with resonant parameter. Take super-Ohmic for example, it oscillates very quickly for $p\gg1$ region, but no oscillations are present for $p\ll1$. On the other hand, for the initial state $|\Psi\rangle_2=|e_1e_2\rangle$, the quantum discord increases to a fixed maximum value, then decreases gradually without any oscillations, as shown in fig.~\ref{fig3} (d) $\sim$ (f). For this kind of initial state, the functions of resonant parameters depend on reservoir spectra. For sub-Ohmic or Ohmic reservoir, resonant parameters in $p\gg1$ case  will shorten the life of quantum discord, while in $p\ll1$ case they prolong the life of quantum discord. Instead, adverse results will been observed for super-Ohmic reservoir.

In conclusion, the non-Markovian master equations are derived to study the decoherence dynamics of two qubits weakly coupled to a common and two separate bosonic thermal baths, respectively. We compare the dynamics at zero temperature and at high temperature under different parameters, such as resonant parameters and reservoir spectra, and we find that there are both similarties and differences between the decoherence processes at these two extreme temperatures: firsly, coherence of two qubits interacting with a common reservoir can be preserved longer than that with two independ reservoirs at high temperature, while the reverse result will be seen at zero temperature; secondly, on the one hand for the super-Ohmic reservoir and sub-Ohmic reservoir the discord decay in a very similar manner and both are faster than Ohmic case in short-time region at high temperature, on the other hand the discord decay in a similar manner for sub-Ohmic and Ohmic reservoir and both are slower than super-Ohmic reservoir case in short-time region at zero temperature; thirdly, the difference between the decoherence processes in resonant and off-resonant region at high temperature is similar to that at zero temperature; finally, non-classical correlation of two qubits coupled to a common reservoir can be generated at zero temperature. The results in this paper not only have fundamental interest for the study of quantum discord, but also can be utilized to design quantum computation in the future.

\acknowledgments

This work is supported by the National Natural Science
Foundation of China under Grant No.60977042 and the
Guangdong Natural Science Foundation under Grant
No. 9151027501000070.

\end{document}


\title{The supplementary material\thanks{The supplementary material is related to the
manuscript "Quantum discord dynamics in structured reservoirs", which is
submitted to \textit{Europhysics Letters}.}}
\author{Z. -K. Su\thanks{E-mail:suzhikun@163.com} and S. -J.
Jiang\thanks{E-mail:stsjsj@mail.sysu.edu.cn}\\State Key Laboratory of Optoelectronic Materials and \\Technologies, Sun Yat-Sen University, Guangzhou 510275, China }
\date{2011-5-23}
\maketitle

The solutions to the master equations will be given in this supplementary
material. We aim to demonstrate that the case of independent reservoirs is
very different from the case of a common reservoir with the appearance of
superoperators $J_{1,2,\pm}$.

\section{The master equations}

\bigskip The non-Markovian master equation can be obtained by The
non-Markovian master equation for two qubits coupled to two independent
reservoirs reads%

\begin{align}
\frac{d\rho}{dt}  & =-4\kappa_{1}\rho-i\kappa_{2}J_{0}\rho\nonumber\\
& +2(\kappa_{1}+\mu_{1})K_{-}\rho+2(\kappa_{1}-\mu_{1})K_{+}\rho-4\mu_{1}%
K_{0}\rho,\label{independent me}%
\end{align}
and the master equation for two qubits coupled to a common reservoir takes the form%

\begin{align}
\frac{d\rho}{dt}  & =-4\kappa_{1}\rho-2\kappa_{1}J_{1}\rho-i2\mu_{2}J_{2}%
\rho\nonumber\\
& +2(\kappa_{1}+\mu_{1})J_{-}\rho+2(\kappa_{1}-\mu_{1})J_{+}\rho-i\kappa
_{2}J_{0}\rho\nonumber\\
& +2(\kappa_{1}+\mu_{1})K_{-}\rho+2(\kappa_{1}-\mu_{1})K_{+}\rho-4\mu_{1}%
K_{0}\rho,\label{common me}%
\end{align}
where $J_{0,1,2,\pm}$ and $K_{0,\pm}$ are superoperators defined as\bigskip%
\begin{equation}
K_{-}\rho=\sigma_{-}^{(1)}\rho\sigma_{+}^{(1)}+\sigma_{-}^{(2)}\rho\sigma
_{+}^{(2)},\label{definition1}%
\end{equation}

\begin{equation}
K_{+}\rho=\sigma_{+}^{(1)}\rho\sigma_{-}^{(1)}+\sigma_{+}^{(2)}\rho\sigma
_{-}^{(2)},\label{definition2}%
\end{equation}

\begin{equation}
K_{0}\rho=\frac{\sigma_{+}^{(1)}\sigma_{-}^{(1)}\rho+\rho\sigma_{+}%
^{(1)}\sigma_{-}^{(1)}-\rho}{2}+\frac{\sigma_{+}^{(2)}\sigma_{-}^{(2)}%
\rho+\rho\sigma_{+}^{(2)}\sigma_{-}^{(2)}-\rho}{2},\label{definition3}%
\end{equation}

\begin{equation}
J_{0}\rho=\sigma_{z}^{(1)}\rho+\sigma_{z}^{(2)}\rho-\rho\sigma_{z}^{(1)}%
-\rho\sigma_{z}^{(2)},\label{definition4}%
\end{equation}

\begin{equation}
J_{1}\rho=\sigma_{-}^{(1)}\sigma_{+}^{(2)}\rho+\sigma_{+}^{(1)}\sigma
_{-}^{(2)}\rho+\rho\sigma_{-}^{(1)}\sigma_{+}^{(2)}+\rho\sigma_{+}^{(1)}%
\sigma_{-}^{(2)},\label{definition5}%
\end{equation}

\begin{equation}
J_{2}\rho=\sigma_{+}^{(1)}\sigma_{-}^{(2)}\rho+\sigma_{+}^{(2)}\sigma
_{-}^{(1)}\rho-\rho\sigma_{+}^{(1)}\sigma_{-}^{(2)}-\rho\sigma_{+}^{(2)}%
\sigma_{-}^{(1)},\label{definition6}%
\end{equation}

\begin{equation}
J_{+}\rho=\sigma_{+}^{(1)}\rho\sigma_{-}^{(2)}+\sigma_{+}^{(2)}\rho\sigma
_{-}^{(1)}.\label{definition8}%
\end{equation}

\section{The solutions}

\subsection{For the case of separate reservoirs}

\bigskip

The solution to master eq.(\ref{independent me}) can be obtained with the
algebaraic approach\cite{algebraic} because the superoperators herein satisfy
SU(2) Lie algebraic communication relations, i.e.%

\[
\lbrack J_{0},K_{\pm,0}]=0,
\]

\begin{equation}
\lbrack K_{0},K_{\pm}]=\pm K_{\pm},[K_{-},K_{+}]=-2K_{0}%
.\label{communication relation1}%
\end{equation}

By directly integrating eq.(\ref{independent me}), the formal solution is
obtained as%

\begin{equation}
\rho(t)=e^{-\Gamma}e^{j_{0}J_{0}}\widehat{T}e^{\int\nolimits_{0}^{t}dt(\nu
_{0}K_{0}+\nu_{+}K_{+}+\nu_{-}K_{-})}\rho(0),\label{sol ind me}%
\end{equation}
\ \ where $\widehat{T}$ is the time ordering operator,%

\[
j_{0}=-2i\int\nolimits_{0}^{t}dt\kappa_{2}(\kappa_{2}+\mu_{2}),\nu_{0}%
=-4\mu_{1},
\]

\begin{equation}
\nu_{+}=2(\kappa_{1}-\mu_{1}),\nu_{-}=2(\kappa_{1}+\mu_{1}),\Gamma
=4\int\nolimits_{0}^{t}dt\kappa_{1}.\label{coeffi}%
\end{equation}

The exponential function of superoperator in eq. (\ref{sol ind me}) can be
factorized in the form%

\begin{equation}
\widehat{T}e^{\int\nolimits_{0}^{t}dt(\nu_{0}K_{0}+\nu_{+}K_{+}+\nu_{-}K_{-}%
)}=e^{k_{+}K_{+}}e^{k_{0}K_{0}}e^{k_{-}K_{-}},\label{sol}%
\end{equation}
where $k_{0},k_{+}$ and $k_{-}$ satisfy the following differential equations%

\begin{equation}
\overset{.}{k}_{+}=\nu_{+}-\nu_{-}k_{+}^{2}+\nu_{0}k_{+},\label{diff1}%
\end{equation}

\begin{equation}
\overset{.}{k}_{0}=\nu_{0}-2\nu_{-}k_{+},\label{diff2}%
\end{equation}

\begin{equation}
\overset{.}{k}_{-}=\nu_{-}\exp(k_{0}),\label{diff3}%
\end{equation}
and using the following relations%

\begin{align}
e^{j_{0}J_{0}}  & =(ch\frac{j_{0}}{4}+\sigma_{z}^{(2)}sh\frac{j_{0}}%
{4})(ch\frac{j_{0}}{4}+\sigma_{z}^{(1)}sh\frac{j_{0}}{4})\nonumber\\
& \cdot\rho\cdot(ch\frac{j_{0}}{4}+\sigma_{z}^{(1)}sh\frac{j_{0}}{4}%
)(ch\frac{j_{0}}{4}+\sigma_{z}^{(2)}sh\frac{j_{0}}{4}),\label{sol 1}%
\end{align}%
\begin{equation}
e^{k_{+}K_{+}}\rho=\rho+j_{+}(\sigma_{+}^{(1)}\rho\sigma_{-}^{(1)}+\sigma
_{+}^{(2)}\rho\sigma_{-}^{(2)})+(j_{+})^{2}\sigma_{+}^{(1)}\sigma_{+}%
^{(2)}\rho\sigma_{-}^{(1)}\sigma_{-}^{(2)},\label{sol 2}%
\end{equation}

\begin{equation}
e^{k_{-}K_{-}}\rho=\rho+j_{-}(\sigma_{-}^{(1)}\rho\sigma_{+}^{(1)}+\sigma
_{-}^{(2)}\rho\sigma_{+}^{(2)})+(j_{-})^{2}\sigma_{-}^{(1)}\sigma_{-}%
^{(2)}\rho\sigma_{+}^{(1)}\sigma_{+}^{(2)},\label{sol 3}%
\end{equation}

\begin{align}
e^{k_{0}K_{0}}\rho & =e^{-\frac{k_{0}}{2}}[1+(e^{\frac{k_{0}}{2}}-1)\sigma
_{+}^{(2)}\sigma_{-}^{(2)}]\{e^{-\frac{k_{0}}{2}}[1+(e^{\frac{k_{0}}{2}%
}-1)\sigma_{+}^{(1)}\sigma_{-}^{(1)}]\nonumber\\
& \cdot\rho\cdot\lbrack1+(e^{\frac{k_{0}}{2}}-1)\sigma_{+}^{(1)}\sigma
_{-}^{(1)}]\}[1+(e^{\frac{k_{0}}{2}}-1)\sigma_{+}^{(2)}\sigma_{-}%
^{(2)}].\label{sol 4}%
\end{align}

Therefore, one can obtain%

\begin{equation}
\rho(t)=\left(
\begin{array}
[c]{cccc}%
\rho_{11}(t) & \rho_{12}(t) & \rho_{13}(t) & \rho_{14}(t)\\
\rho_{21}(t) & \rho_{22}(t) & \rho_{23}(t) & \rho_{24}(t)\\
\rho_{31}(t) & \rho_{32}(t) & \rho_{33}(t) & \rho_{34}(t)\\
\rho_{41}(t) & \rho_{42}(t) & \rho_{43}(t) & \rho_{44}(t)
\end{array}
\right)  ,\label{matrix for in}%
\end{equation}
with%

\begin{align*}
\rho_{11}(t)  & =(e^{k_{0}}+2k_{+}k_{-}+e^{-k_{0}}k_{+}^{2}k_{-}^{2})\rho
_{11}(0)+(k_{+}+e^{-k_{0}}k_{+}^{2}k_{-})\rho_{22}(0)\\
& +(k_{+}+e^{-k_{0}}k_{+}^{2}k_{-})\rho_{33}(0)+e^{-k_{0}}k_{+}^{2}\rho
_{44}(0),
\end{align*}

\begin{align*}
\rho_{22}(t)  & =(k_{-}+e^{-k_{0}}k_{+}k_{-}^{2})\rho_{11}(0)+(1+e^{-k_{0}%
}k_{+}k_{-})\rho_{22}(0)\\
& +(1+e^{-k_{0}}k_{+}k_{-})\rho_{33}(0)+e^{-k_{0}}k_{+}\rho_{44}(0),
\end{align*}

\begin{align*}
\rho_{33}(t)  & =(k_{-}+e^{-k_{0}}k_{+}k_{-}^{2})\rho_{11}(0)+e^{-k_{0}}%
k_{+}k_{-}\rho_{22}(0)\\
& +e^{-k_{0}}k_{+}k_{-}\rho_{33}(0)+e^{-k_{0}}k_{+}\rho_{44}(0),
\end{align*}
\ \ \ %

\[
\rho_{44}(t)=e^{-k_{0}}k_{-}^{2}\rho_{11}(0)+e^{-k_{0}}k_{-}[\rho_{22}%
(0)+\rho_{33}(0)]++e^{-k_{0}}\rho_{44}(0),
\]
\ \ \ %

\[
\rho_{21}(t)=(e^{k_{0}/2}+e^{-k_{0}/2}k_{+}k_{-})\rho_{21}(0)+e^{-k_{0}%
/2}k_{+}\rho_{43}(0),
\]

\[
\rho_{31}(t)=(e^{k_{0}/2}+e^{-k_{0}/2}k_{+}k_{-})\rho_{31}(0)+e^{-k_{0}%
/2}k_{+}\rho_{42}(0),
\]

\[
\rho_{12}(t)=(e^{k_{0}/2}+e^{-k_{0}/2}k_{+}k_{-})\rho_{12}(0)+e^{-k_{0}%
/2}k_{+}\rho_{34}(0),
\]

\[
\rho_{42}(t)=e^{-k_{0}/2}k_{-}\rho_{31}(0)+e^{-k_{0}/2}\rho_{42}(0),
\]

\[
\rho_{13}(t)=(e^{k_{0}/2}+e^{-k_{0}/2}k_{+}k_{-})\rho_{13}(0)+e^{-k_{0}%
/2}k_{+}\rho_{24}(0),
\]

\[
\rho_{43}(t)=e^{-k_{0}/2}k_{-}\rho_{21}(0)+e^{-k_{0}/2}\rho_{43}(0),
\]

\[
\rho_{24}(t)=e^{-k_{0}/2}k_{-}\rho_{13}(0)+e^{-k_{0}/2}\rho_{24}(0),
\]

\[
\rho_{34}(t)=e^{-k_{0}/2}k_{-}\rho_{12}(0)+e^{-k_{0}/2}\rho_{34}(0),
\]

\begin{equation}
\rho_{14}(t)=\rho_{14}(0),\rho_{23}(t)=\rho_{23}(0),\rho_{32}(t)=\rho
_{32}(0),\rho_{41}(t)=\rho_{41}(0).\label{element for in}%
\end{equation}
From eq.(\ref{element for in}), we can see that $J_{0},K_{\pm,0}$ can't affect
the anti-diagonal elements of the density matrix. In the next section, we aim
to find out wheather the anti-diagonal elements of the density matrix will be
affected in the case of a common reservoir.

\ \ \ 

\subsection{For the case of a common reservoir}

\bigskip The formal solution to eq.(\ref{common me}) is obtained as%

\begin{equation}
\rho(t)=e^{-\Gamma}e^{j_{0}J_{0}}\widehat{T}e^{\int\nolimits_{0}%
^{t}dt(\varepsilon_{1}J_{1}+\varepsilon_{2}J_{2}+\varepsilon_{+}%
J_{+}+\varepsilon_{-}J_{-}+\nu_{0}K_{0}+\nu_{+}K_{+}+\nu_{-}K_{-})}%
\rho(0).\label{sol to com}%
\end{equation}

The superoperators satisfy the communication relations%

\[
\lbrack J_{0},J_{\pm,1,2}]=0,
\]

\[
\lbrack J_{0},K_{\pm,0}]=0,
\]

\[
\lbrack J_{1},K_{-}]\rho=\sigma_{-}^{(1)}\sigma_{z}^{(2)}\rho\sigma_{+}%
^{(2)}+\sigma_{-}^{(2)}\sigma_{z}^{(1)}\rho\sigma_{+}^{(1)}+\sigma_{-}%
^{(1)}\rho\sigma_{z}^{(1)}\sigma_{+}^{(2)}+\sigma_{-}^{(2)}\rho\sigma
_{z}^{(2)}\sigma_{+}^{(1)},
\]

\[
\lbrack J_{1},K_{+}]\rho=-\sigma_{+}^{(1)}\sigma_{z}^{(2)}\rho\sigma_{-}%
^{(2)}-\sigma_{+}^{(2)}\sigma_{z}^{(1)}\rho\sigma_{-}^{(1)}-\sigma_{+}%
^{(1)}\rho\sigma_{z}^{(1)}\sigma_{-}^{(2)}-\sigma_{+}^{(2)}\rho\sigma
_{z}^{(2)}\sigma_{-}^{(1)},
\]

\begin{align*}
\lbrack J_{1},K_{0}]\rho & =\frac{1}{2}\{\sigma_{+}^{(2)}\sigma_{-}%
^{(1)}\sigma_{+}^{(1)}\sigma_{-}^{(1)}\rho+\sigma_{+}^{(1)}\sigma_{-}%
^{(2)}\sigma_{+}^{(2)}\sigma_{-}^{(2)}\rho+\rho\sigma_{+}^{(2)}\sigma
_{-}^{(2)}\sigma_{+}^{(2)}\sigma_{-}^{(1)}+\rho\sigma_{+}^{(1)}\sigma
_{-}^{(1)}\sigma_{+}^{(1)}\sigma_{-}^{(2)},\\
& -[\sigma_{-}^{(2)}\sigma_{+}^{(1)}\sigma_{-}^{(1)}\sigma_{+}^{(1)}\rho
+\rho\sigma_{-}^{(1)}\sigma_{+}^{(1)}\sigma_{-}^{(1)}\sigma_{+}^{(2)}%
+\sigma_{-}^{(1)}\sigma_{+}^{(2)}\sigma_{-}^{(2)}\sigma_{+}^{(2)}\rho
+\rho\sigma_{-}^{(2)}\sigma_{+}^{(2)}\sigma_{-}^{(2)}\sigma_{+}^{(1)}]\},
\end{align*}

\[
\lbrack J_{1},J_{2}]=0,
\]

\[
\lbrack J_{1},J_{-}]\rho=\sigma_{z}^{(2)}\sigma_{-}^{(1)}\rho\sigma_{+}%
^{(1)}+\sigma_{z}^{(1)}\sigma_{-}^{(2)}\rho\sigma_{+}^{(2)}+\sigma_{-}%
^{(2)}\rho\sigma_{+}^{(2)}\sigma_{z}^{(1)}+\sigma_{-}^{(1)}\rho\sigma
_{+}^{(1)}\sigma_{z}^{(2)},
\]

\[
\lbrack J_{1},J_{+}]\rho=-\sigma_{z}^{(1)}\sigma_{+}^{(2)}\rho\sigma_{-}%
^{(2)}-\sigma_{z}^{(2)}\sigma_{+}^{(1)}\rho\sigma_{-}^{(1)}-\sigma_{+}%
^{(1)}\rho\sigma_{-}^{(1)}\sigma_{z}^{(2)}-\sigma_{+}^{(2)}\rho\sigma
_{-}^{(2)}\sigma_{z}^{(1)},
\]

\[
\lbrack J_{2},J_{-}]\rho=\sigma_{z}^{(1)}\sigma_{-}^{(2)}\rho\sigma_{+}%
^{(2)}+\sigma_{z}^{(2)}\sigma_{-}^{(1)}\rho\sigma_{+}^{(1)}-\sigma_{-}%
^{(1)}\rho\sigma_{+}^{(1)}\sigma_{z}^{(2)}-\sigma_{-}^{(2)}\rho\sigma
_{+}^{(2)}\sigma_{z}^{(1)},
\]

\[
\lbrack J_{2},J_{+}]\rho=-\sigma_{z}^{(2)}\sigma_{+}^{(1)}\rho\sigma_{-}%
^{(1)}-\sigma_{z}^{(1)}\sigma_{+}^{(2)}\rho\sigma_{-}^{(2)}+\sigma_{+}%
^{(2)}\rho\sigma_{-}^{(2)}\sigma_{z}^{(1)}+\sigma_{+}^{(2)}\rho\sigma
_{+}^{(2)}\sigma_{z}^{(1)},
\]

\[
\lbrack J_{2},K_{-}]\rho=\sigma_{-}^{(2)}\sigma_{z}^{(1)}\rho\sigma_{+}%
^{(1)}+\sigma_{-}^{(1)}\sigma_{z}^{(2)}\rho\sigma_{+}^{(2)}-\sigma_{-}%
^{(2)}\rho\sigma_{z}^{(2)}\sigma_{+}^{(1)}-\sigma_{-}^{(1)}\rho\sigma
_{z}^{(1)}\sigma_{+}^{(2)},
\]

\[
\lbrack J_{2},K_{+}]\rho=-\sigma_{+}^{(1)}\sigma_{z}^{(2)}\rho\sigma_{-}%
^{(2)}-\sigma_{+}^{(2)}\sigma_{z}^{(1)}\rho\sigma_{-}^{(1)}+\sigma_{+}%
^{(1)}\rho\sigma_{z}^{(1)}\sigma_{-}^{(2)}+\sigma_{+}^{(2)}\rho\sigma
_{z}^{(2)}\sigma_{-}^{(1)},
\]

\begin{align*}
\lbrack J_{2},K_{0}]\rho & =\frac{1}{2}(\sigma_{+}^{(1)}\sigma_{-}^{(2)}%
\sigma_{+}^{(2)}\sigma_{-}^{(2)}\rho+\sigma_{+}^{(2)}\sigma_{-}^{(1)}%
\sigma_{+}^{(1)}\sigma_{-}^{(1)}\rho-\rho\sigma_{+}^{(1)}\sigma_{-}%
^{(1)}\sigma_{+}^{(1)}\sigma_{-}^{(2)}-\rho\sigma_{+}^{(2)}\sigma_{-}%
^{(2)}\sigma_{+}^{(2)}\sigma_{-}^{(1)})\\
& -\frac{1}{2}(\sigma_{-}^{(2)}\sigma_{+}^{(1)}\sigma_{-}^{(1)}\sigma
_{+}^{(1)}\rho-\rho\sigma_{-}^{(1)}\sigma_{+}^{(1)}\sigma_{-}^{(1)}\sigma
_{+}^{(2)}+\sigma_{-}^{(1)}\sigma_{+}^{(2)}\sigma_{-}^{(2)}\sigma_{+}%
^{(2)}\rho-\rho\sigma_{-}^{(2)}\sigma_{+}^{(2)}\sigma_{-}^{(2)}\sigma
_{+}^{(1)}),
\end{align*}

\[
\lbrack J_{-},J_{+}]\rho=-\sigma_{z}^{(1)}\rho-\rho\sigma_{z}^{(1)}-\rho
\sigma_{z}^{(2)}-\sigma_{z}^{(2)}\rho,
\]

\[
\lbrack J_{-},K_{+}]\rho=-\sigma_{z}^{(1)}\rho\sigma_{-}^{(1)}\sigma_{+}%
^{(2)}-\sigma_{-}^{(1)}\sigma_{+}^{(2)}\rho\sigma_{z}^{(2)}-\sigma_{-}%
^{(2)}\sigma_{+}^{(1)}\rho\sigma_{z}^{(1)}-\sigma_{z}^{(2)}\rho\sigma
_{-}^{(2)}\sigma_{+}^{(1)},
\]

\[
\lbrack J_{-},K_{-}]\rho=0,
\]

\[
\lbrack J_{-},K_{0}]\rho=\frac{1}{2}(-\sigma_{z}^{(1)}\sigma_{-}^{(1)}%
\rho\sigma_{+}^{(2)}-\sigma_{-}^{(1)}\rho\sigma_{+}^{(2)}\sigma_{z}%
^{(2)}-\sigma_{-}^{(2)}\rho\sigma_{+}^{(1)}\sigma_{z}^{(1)}-\sigma_{z}%
^{(2)}\sigma_{-}^{(2)}\rho\sigma_{+}^{(1)}),
\]

\[
\lbrack J_{+},K_{-}]\rho=\sigma_{z}^{(1)}\rho\sigma_{+}^{(1)}\sigma_{-}%
^{(2)}+\sigma_{+}^{(1)}\sigma_{-}^{(2)}\rho\sigma_{z}^{(2)}+\sigma_{+}%
^{(2)}\sigma_{-}^{(1)}\rho\sigma_{z}^{(1)}+\sigma_{z}^{(2)}\rho\sigma
_{+}^{(2)}\sigma_{-}^{(1)},
\]

\[
\lbrack J_{+},K_{+}]\rho=0,
\]
and%
\begin{equation}
\lbrack J_{+},K_{0}]\rho=-\frac{1}{2}(\sigma_{+}^{(1)}\sigma_{-}^{(1)}%
\sigma_{+}^{(1)}\rho\sigma_{-}^{(2)}+\sigma_{+}^{(2)}\rho\sigma_{-}%
^{(1)}\sigma_{+}^{(1)}\sigma_{-}^{(1)}+\sigma_{+}^{(2)}\sigma_{-}^{(2)}%
\sigma_{+}^{(2)}\rho\sigma_{-}^{(1)}+\sigma_{+}^{(1)}\rho\sigma_{-}%
^{(2)}\sigma_{+}^{(2)}\sigma_{-}^{(2)}).\label{communication relation2}%
\end{equation}

We then approximate\cite{approximant} the exponential in the right-hand side
of eq.(\ref{sol to com}). The first-order approximant is given by%

\begin{align}
& \widehat{T}e^{\int\nolimits_{0}^{t}dt(\varepsilon_{1}J_{1}+\varepsilon
_{2}J_{2}+\varepsilon_{+}J_{+}+\varepsilon_{-}J_{-}+\nu_{0}K_{0}+\nu_{+}%
K_{+}+\nu_{-}K_{-})}\nonumber\\
& =e^{-it(\varepsilon_{1}J_{1}+\varepsilon_{2}J_{2}+\varepsilon_{+}%
J_{+}+\varepsilon_{-}J_{-})}e^{-it(\nu_{0}K_{0}+\nu_{+}K_{+}+\nu_{-}K_{-}%
)}.\label{exponential com}%
\end{align}

Solving eq.(\ref{sol to com}) even for its first-order approximant is a non
trivial task, but we can see from eq.(\ref{exponential com}) that the
superoperators $J_{1,2,\pm}$ will affect on the initial state of the system
and further caculation shows that $J_{1,2,\pm}$\ will affect on all
elements,\ including anti-diagonal elements, of the density matrix. It should
be pointed out that we evaluate dynamics of qubits coupled to a common
reservoir numerically in the manuscript. We are intertested in what will be
induced by the difference between the case of independent reservoirs and the
case of a common reservoir and whether it will vary from high-temperature
region to zero-temperature region. \ \